\documentclass[10pt,conference,twocolumn]{IEEEtran}
 
\usepackage[nolist]{acronym}
\makeatother
\IEEEoverridecommandlockouts

\usepackage{listings}
\usepackage{amsfonts,amssymb,stmaryrd,dsfont,calrsfs} 
\usepackage[fleqn]{amsmath}
\usepackage[usenames,dvipsnames]{xcolor}
\usepackage{cite,booktabs,multirow}
\usepackage{graphicx}
\usepackage{times}
\usepackage{array}

\usepackage{listings}

\usepackage[hidelinks]{hyperref}
\usepackage{epstopdf}
\usepackage[numbers]{natbib}
\usepackage{lettrine}
\usepackage{tikz}
 
\usepackage{tikzscale}
\usepackage[footnote, nomargin, draft]{fixme}
\usepackage[para]{threeparttable}

\def\ninept{\def\baselinestretch{.95}\let\normalsize\small\normalsize}
\author{}
\author{
\IEEEauthorblockN{Egon Kidmose, Emad Ebeid, Rune Hylsberg Jacobsen }
\IEEEauthorblockA{Department of Engineering, Aarhus University, Denmark\\
  }
}

\title{ A Framework for Detecting and Translating User Behavior from Smart Meter Data
}

\acrodef{DSO}{Distribution System Operator}
\acrodef{NALM}{Nonintrusive Appliance Load Monitoring}
\acrodef{REST}{REpresentational State Transfer}
\acrodef{UML}{Unified Modeling Language}
\acrodef{OMG}{Object Management Group}
\acrodef{MTL}{Model-to-Text Language}
\acrodef{HTTP}{HyperText Transfer Protocol}
\acrodef{IP}{Internet Protocol}
\acrodef{IPv6}{IP version 6}
\acrodef{LDA}{Load Disaggregation Algorithm}
\acrodef{SVM}{Support Vector Machine}
\acrodef{TNR}{True Negative Rate}
\acrodef{TPR}{True Positive Rate}
\acrodef{TP}{True Positive}
\acrodef{TN}{True Negative}
\acrodef{FP}{False Positive}
\acrodef{FN}{False NEgative}
\acrodef{NLP}{Natural Language Processing}
\acrodef{ML}{Machine-Learning}
\acrodef{DBA}{DataBase and Analytics}

\begin{document}
\maketitle
\sloppy

\begin{abstract}

The European adoption of smart electricity meters triggers the developments of new value-added service for smart energy and optimal consumption. Recently, several algorithms and tools have been built to analyze smart meter's data.    
This paper introduces an open framework and prototypes for detecting and presenting user behavior from its smart meter power consumption data. The framework aims at presenting the detected user behavior in natural language reports. In order to validate the proposed framework, an experiment has been performed and the results have been presented. 

\end{abstract}


\section{Introduction}
\label{sec:introduction} 
In a recent report, i.e., ``Benchmarking smart metering deployment in the EU-27 with a focus on electricity'', the Commission has accelerated smart meter deployment in European households. 
A roll-out target of 80\% market penetration of smart electricity meters by 2020 has been required by Member States given that a long-term cost-benefit analysis proves to be positive.
Besides being essential for electricity billing, smart meters have been used as a vehicle for delivering value-added services such as
providing \acp{DSO} with diagnostic information about the distribution grid~\cite{Depuru20112736} or by permitting third-party application providers to deliver grid-related information services to residential homes~\cite{jacobsen2014}.
Furthermore, the higher temporal resolution of consumption data, offered by smart meters, allows physical events resulting from user behavior to be detected and analyzed by using 
a \ac{NALM} system~\cite{Weiss2012}.

\begin{figure}[t]
\centering

\includegraphics[width=0.78\linewidth]{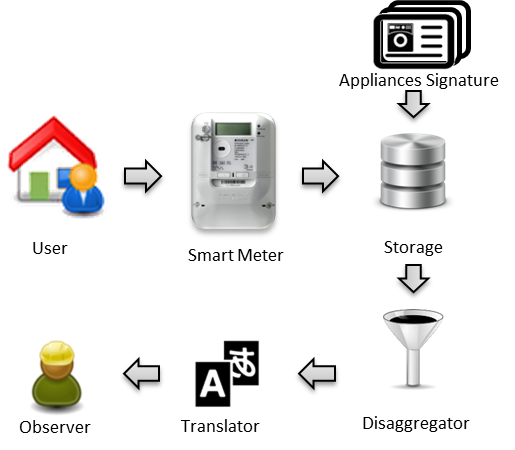}
\center \caption{ Overview of the data flow architecture}

\label{fig:overview}
\end{figure}

This paper rests on recent advancements in \ac{NALM} based on smart meter data readings with high temporal resolution. 
A conceptual framework (shown in Figure~\ref{fig:overview}) for detecting user behavior from the electricity usage fingerprints, resulting from activities in the residential home, is proposed. 
The key novelty of the research comes from a combination of simple \ac{ML}  techniques for event recognition with a subsequent analysis and translation of information into natural language.
After an initial training period, the responses to the actual user behavior can be delivered in near real-time. 
The immediate benefit of the proposed framework stems from the fact that an observer no longer needs to be a skilled technician but instead can rely on comprehensible reports on user behavior. 

A domain where the framework can be applied is elderly care, where a report in natural language will enable caretakers to take the role of the observer. 
By cutting away transportation overhead, this has the potential to allow caretakers to spend more time paying attention to indicators of discomfort or worse.
The paper is structured as follows; 
Section~\ref{sec:soa} gives a brief overview of the related work. 
Section~\ref{sec:bg} shows the techniques and modeling languages used in this work. 
Section~\ref{sec:fw} presents the proposed framework and Section~\ref{sec:results} demonstrates the applicability of the framework via a test case. 
Section~\ref{sec:conclusions} draws the conclusion and outlines future work.  

\section{State of the Art}
\label{sec:soa} 
While frameworks for \ac{NALM} by smart meter power consumption data forms a relatively new research field, diverse algorithms and tools have been presented to implement these frameworks. 
In~\cite{Weiss2012}, authors present an infrastructure and a specialized algorithm that provide users with real-time
feedback on their electricity consumption. 
They achieve $86.8\%$ accuracy in detecting ON/OFF switching events. 
In~\cite{Ruzzelli2010}, Ruzzelli et al. present a smart system for recognition of electrical appliance activities in real-time. 
Their system provides $84.6\%$ accuracy in determining the set of appliances being used. 

In the same context, the work reported in \cite{rune2014} defines a service-oriented architecture for the collection of electricity data from resource constrained devices in residential homes. 
In the same work, the \ac{REST} principle is applied when designing the application layer protocols and a database cloud service, providing storage for other elements in the architecture. 

Besides the simplicity stemming from only using smart meter data, research has also focused on the use of data with low time resolution. 
The works of~\cite{Ruzzelli2010} and ~\cite{liao2014non} both rely on data collected at $\frac{1}{60}$ Hz, where the latter achieve a precision of $76.1\%$ in detecting switching events.

Regarding the representation of the analyzed smart meter data results; generation of natural language from software models is considered as a key target in this point. Burden et al. in~\cite{Burden:2011}, investigate the possibilities to generate natural language text from \ac{UML} diagrams. 
They use a static diagram (i.e., class diagram) transformed into an intermediate linguistic model to demonstrate their approach. 
They show that the generated texts are grammatically correct. In this work, we have followed the same approach to generate natural language reports from high-level models.
 
\section{Background}
\label{sec:bg} 
This section gives a brief overview of appliances detecting algorithms,  natural language,' and interfaces that are used to exchange data between the framework elements.

\subsection{Load Disaggregation Algorithm}
Defined as an algorithm that takes data on aggregated electricity loads from multiple appliances, as input, and outputs disaggregated loads for individual appliances~\cite{witten2005data}. 
Combined with a non-intrusive approach to obtain the data, it forms a method for \ac{NALM}. 
Assuming that labels with information about appliance load is available for some of the load data, the problem of disaggregation is similar to a supervised learning problem known from \ac{ML} or a problem of statistical regression~\cite{witten2005data}. 
Another problem, related to load disaggregation is that of detecting event states, typically ``ON'' or ``OFF''. 

\subsection{Model-to-Text}
An Eclipse project is concerned with the generation of textual artifacts from high-level models. 
\ac{OMG} specifies a correlated language named \ac{MTL} to express its transformation. 
Acceleo~\cite{acceleo}, an Eclipse plugin tool, is a pragmatic implementation of the \ac{MTL} standard. It is widely used by software engineers to generate code from high-level models.

\subsection{Interfaces}
\ac{REST} is a design style for designing application based on web services, calling for simple (client-server, stateless, self-documenting) interfaces building on \ac{HTTP}.
The openness, modularity, interoperability, and security provided by \ac{REST} are beneficial when designing interfaces for an open framework such as the proposed. 
The \ac{IP} suite is a key building block for cloud services, due to its widespread use. 
As discussed in~\cite{rune2014}, the larger address range of \ac{IPv6} is necessary to assign each device an address and thereby observe the End-to-End principle. 

ZigBee is implemented on the top of IEEE 802.15.4 standard and widely used in home automation applications, offering low transmission rates over a low-power wireless radio links.
 
\section{Proposed Framework}
\label{sec:fw} 
In this section, the context of the framework is outlined and elements of the framework are defined. 

The context is that of a \textit{user} living in his/her home. 
In the home, some electrical \textit{appliances} are installed and when a user makes use of an appliance it is referred to as a \textit{usage}. 
The framework interfaces with user consumption data through a \textit{smart meter} by receiving data on electrical power consumption. 
The framework is then responsible for establishing the user's usages of appliances and outputs the result as natural language report. 
An \textit{observer} will receive this natural language and thereby obtain information about the user behavior without having to be physically present. 

\begin{figure}[h]
  \centering
\includegraphics[width=0.9\linewidth]{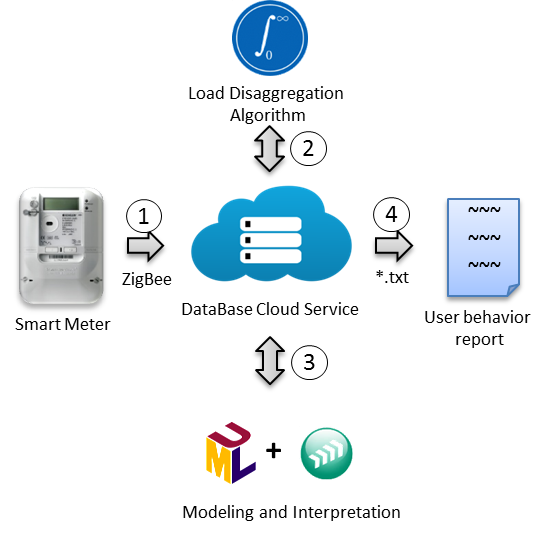}
  \caption{Tool-chain structure of the proposed framework}
  \label{fig:toolchain}
\end{figure}

The framework is outlined in Figure~\ref{fig:toolchain} with its five components; numerical labels are used to denote the data flow directions.
A \textit{smart meter} will be responsible for collecting consumption data of the home. 
By using ZigBee communication data will be transferred to a DataBase Cloud Service (label~1) and a gateway has to be deployed in the home as discussed in \cite{rune2014}. 
The \textit{DataBase Cloud Service} provides a data storage for the resource-constrained smart meters. 
With \ac{REST}ful web interface the service has potential for great scalability, like many other cloud services, while still offering constrained devices a simple interface. 
Besides providing a way to how to store potentially data, which is not discussed in other work on \ac{NALM}, this framework describes a component which can be expected to scale well and has the benefits of consolidating storage. 
To make use of the collected data, it is proposed to deploy a \textit{\ac{LDA}}~\cite{Marceau20001389} (label~2) as a cloud service. The LDA service takes measurement data from the DataBase Cloud Service as input and disaggregates the aggregated electricity consumption to produces data on appliance usage. 
Multiple implementations of LDAs exist as discussed in Section \ref{sec:soa} and one of the key features of this framework is the capability to use various algorithms without rebuilding other components~\cite{jacobsen2014}. 
This is a benefit in both evaluation and deployment as it simplifies comparison and replacement of the \ac{LDA}. 
As a novel idea in this context, {\ac{NLP} is applied to present information derived from electricity usage data to the receiver as natural language (label~3). 
Depending on the specific requirements this can be audio or text (label~4). 
An important feature is that the \ac{NLP} components can be replaced without affecting other components in the framework. 
Figure~\ref{fig:model} shows a part of a developed UML class diagram of the framework architecture. The class diagram depicts the main used classes with their attributes and relationships with each other. In this way, objects of those classes can easily be instantiated and information related to each object can be tagged. Afterward, such objects will be transformed into a natural language as it explained in details in~\cite{Burden:2011}. 

\begin{figure}[h!]
\centering
\includegraphics[width=0.85\linewidth]{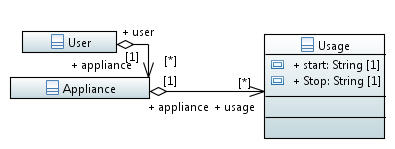}
\caption{Part of the framework architecture that related to the natural language generation}
\label{fig:model}
\end{figure}
 
\section{Experimental results}
\label{sec:results} 
This section demonstrates the applicability of the proposed framework via a test case.
\subsection{Electricity Traces of Appliances}
\label{sec:exp.res.smartmeter}
In place of a real smart meter and the data to be obtained from it, this work in its current state relies on the \textit{Tracebase} dataset provided by~\cite{reinhardt12tracebase}. 
The dataset contains 1836 traces of 24-hours duration, from 159 different appliances of 43 types, with average power consumption, sampled at 1 Hz. 
For collecting, the authors of~\cite{reinhardt12tracebase} used the smart plug product named Circle, by PlugWise~\cite{plugwise10}. 
While providing precise per-appliance measurements, the Circle is an intrusive device, that must be installed between power outlets and appliance power plugs. 

To simulate a smart meter, two days of data from six appliances are considered --- one for test and one for training. 
The test and training sets are constructed by summing across appliances for each time of day, resulting in two virtual days of smart meter measurements.
Any correlation information between appliances, such as the \textit{PC-Desktop} and \textit{Monitor-TFT} tending to be ON at the same time, is lost in the described process. 
Loss of information is expected to make the task of the \ac{LDA} harder, leading to a worse performance in the evaluation and thereby erring on the side of caution.
The labeling with a usage of appliances is done manually by visually inspecting the dataset, introducing a source of error. 
Both errors are to be eliminated in future work by using a smart meter to measure multiple devices and by recording the true appliance usage.

\subsection{Storage Service}
\label{sec:exp.res.storageservice}
Without a deployed smart meter, there is no need for a storage service and it is not yet implemented directly in this project. 
However, in related research efforts, the authors have obtained experience with the development and implementation of the \ac{DBA} service of~\cite{rune2014}, and plan to use it in the near future. 

\subsection{Load Disaggregation Algorithm}
\label{sec:exp.res.nalm}
To verify that the \ac{LDA} component is feasible, two prototypes have been implemented using two different supervised \ac{ML} techniques, namely \ac{SVM} and Random Forest~\cite{witten2005data}. 
For each \ac{LDA}, a classifier for a home is built on the training data, obtained as discussed in Section~\ref{sec:exp.res.smartmeter}. 
To evaluate, the classifier is applied to the test dataset, resulting in predictions of which appliances are in use at each time interval. 
Usage patterns are obtained by sliding a time window and by observing changes in which appliances are in use. 
A change to ON signifies the starting time of a usage and the next change to OFF for the device signifies the stopping time of the same usage.

\ac{SVM} results in an overall accuracy of 94.0\% and F1-score of 77.3\% with worst per appliance F1-score being 38.4\%. 
Random Forrest, on the other, hand provides an overall accuracy of 94.3\% and F1-score of 78.3\%, with worst per appliance F1-score of 39.2\%. 
The difference in performance between the two methods is too small to ascribe any significance. Therefore,
the results for the worst performing \ac{SVM} is considered in the following. 

Table \ref{tab:disagregation.perfomance} shows that the \ac{LDA} is good at determining when appliances are OFF, as the \ac{TNR} is high. It is also good at determining when appliances are ON, except for the Lamp and the TV-LCD, which shows a low \ac{TPR} and thereby a low F1-score. 
The counts of \ac{TP}, \ac{FP}, \ac{TN} and \ac{FN} are available in Table \ref{tab:disagregator.counts}.

\begin{table}[htb]
  \centering
  \caption{\ac{LDA} (\ac{SVM}) test performance metrics.}
  \scalebox{1}{
    \begin{threeparttable}
      \begin{tabular}{lccccc}
        \toprule \addlinespace[2mm]
        & Prec. & Acc. & TPR & TNR & F1 \\ 
        \midrule \addlinespace[2mm]
        TV-CRT         & 0.998 & 0.976 & 0.855 & 1.000 & 0.921 \\ 
        PC-Desktop     & 0.820 & 0.934 & 0.975 & 0.919 & 0.890 \\ 
        Cooking-stove   & 0.932 & 0.997 & 0.997 & 0.997 & 0.964 \\ 
        Lamp           & 0.621 & 0.887 & 0.288 & 0.974 & 0.394 \\ 
        Monitor-TFT    & 0.613 & 0.930 & 0.988 & 0.923 & 0.757 \\ 
        TV-LCD         & 0.571 & 0.934 & 0.402 & 0.976 & 0.471 \\ 
        Overall       & 0.780 & 0.943 & 0.787 & 0.967 & 0.783 \\
        \bottomrule
      \end{tabular}
      \footnotesize
      \begin{tablenotes}
        \item Prec.: $\frac{TP}{TP+FP}$.
        \item Acc.: $\frac{TP+TN}{TP+FP+FN+TN}$.
        \item TPR.: $\frac{TP}{TP+FN}$.
        \item TNR: $\frac{TN}{TN+FP}$.
        \item F1: $\frac{2 TP}{2 TP + FP + FN}$.
      \end{tablenotes}
    \end{threeparttable}
  }

  \label{tab:disagregation.perfomance}
\end{table}

\begin{table}[htb]
  \centering
  \caption{\ac{LDA} (\ac{SVM}) test error counts.}
  \scalebox{1}{
    \begin{threeparttable}
      \begin{tabular}{lcccc}
        \toprule \addlinespace[2mm]
        & TP & FP & TN & FN \\
        \midrule \addlinespace[2mm]
        TV-CRT         & 12048 & 21 & 72280 & 2049 \\ 
        PC-Desktop     & 23159 & 5099 & 57538 & 602 \\ 
        Cooking-stove   & 3107 & 226 & 83057 & 8 \\ 
        Lamp           & 3177 & 1935 & 73428 & 7858 \\ 
        Monitor-TFT    & 9336 & 5890 & 71056 & 116 \\ 
        TV-LCD         & 2541 & 1912 & 78159 & 3786 \\ 
        Overall       & 53368 & 15083 & 435518 & 14419 \\
        \bottomrule
      \end{tabular}
      \footnotesize
      \begin{tablenotes}
        \item TP: True Positives.
        \item FP: False Positives.
        \item TN: True Negatives.
        \item FN: False Negative.
      \end{tablenotes}
    \end{threeparttable}
  }
  \label{tab:disagregator.counts}
\end{table}

\subsection{\acl{NLP}}
\label{sec:exp.res.nlp}
\label{sec:m2t}
Figure~\ref{fig:example} shows a part of the generated object diagram of the previously explained class diagram (Figure~\ref{fig:model}). 
The diagram is automatically built from the output of the \ac{LDA}, by mapping it directly into class instances. 
A Python module outputting XML conforming to the format used by Eclipse \ac{MTL} has been implemented specifically for this purpose. 
The tool generates \textit{.uml} files which are visualized using Papyrus UML editor tool as shown in Figure~\ref{fig:example}.  

\begin{figure}[h]
\centering
\includegraphics[width=0.8\linewidth]{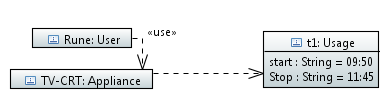}
\caption{Part of the UML model of the case study}
\label{fig:example}
\end{figure}

\subsection{Model-to-Text}
The last step in the synthesis, is to transform \ac{UML} object diagram into a natural language. This step has been done by developing an Acceleo model-to-text generator to parse and convert the model into a natural language. A part of the generator is shown in Figure~\ref{code1}.

\begin{figure}[h]

\lstset{language=Java}          

\begin{lstlisting}  % Start your code-block
 

[let I: Sequence(InstanceSpecification) = model.eAllContents(InstanceSpecification)]
[if (I.classifier->at(i).name = 'User')]
[I->at(i).name/]
[let iUser : Integer = i]
	[for (it : NamedElement | I->at(iUser).clientDependency.supplier)]
	 was using the [it.name/] 
	 [it.clientDependency.supplier.eAllContents(LiteralString).value->sep('from ',' to ','.')/] 
...
\end{lstlisting}
  \caption{A part of the developed Acceleo natural language generator tool }
  \label{code1}
\end{figure}
The automatically generated natural language report from the parsed \ac{UML} object diagram (Figure~\ref{fig:example}) is:
\begin{quote}
\texttt{Rune was using the TV-CRT from 09:50 to 11:45.}
\end{quote} 

\section{Conclusion and Future work}
\label{sec:conclusions} 
A framework for deducing user behavior from smart meter data has been presented. 
Tool-chain structure and prototypes have been described and evaluated for the key components, specifically the \acl{LDA}, the Modeling and the \acl{NLP}. 
The prototypes have been developed to validate the feasibility of the framework. 

Future work includes modeling of the entire framework in details, formalizing interfaces between components and validating them. 
In particular, the issue of multiple users is a topic that has not been discussed in related work. 
Smart meters will be introduced and utilized for acquisition of real and complex electricity consumption data. 
The database cloud service will need to be implemented to support data acquisition, and potentials for sharing data or appliance profiles can be investigated. 
\ac{LDA} performance might be improved through the use of other algorithms or by tuning parameters. 
Establishing the statistical significance will be an important part of the evaluation.

 \section*{Acknowledgment}
 The research leading to these results has received funding from the European Union Seventh Framework Programme (FP7/2007-2013) under {\it grant agreements}
 no.~317761 (SmartHG) and no.~619560 (SEMIAH).
\bibliographystyle{myIEEEtran} 
\bibliography{refs}

\begin{thebibliography}{10}
\providecommand{\url}[1]{#1}
\csname url@samestyle\endcsname
\providecommand{\newblock}{\relax}
\providecommand{\bibinfo}[2]{#2}
\providecommand{\BIBentrySTDinterwordspacing}{\spaceskip=0pt\relax}
\providecommand{\BIBentryALTinterwordstretchfactor}{4}
\providecommand{\BIBentryALTinterwordspacing}{\spaceskip=\fontdimen2\font plus
\BIBentryALTinterwordstretchfactor\fontdimen3\font minus
  \fontdimen4\font\relax}
\providecommand{\BIBforeignlanguage}[2]{{%
\expandafter\ifx\csname l@#1\endcsname\relax
\typeout{** WARNING: IEEEtran.bst: No hyphenation pattern has been}%
\typeout{** loaded for the language `#1'. Using the pattern for}%
\typeout{** the default language instead.}%
\else
\language=\csname l@#1\endcsname
\fi
#2}}
\providecommand{\BIBdecl}{\relax}
\BIBdecl

\bibitem{Depuru20112736}
\BIBentryALTinterwordspacing
S.~S. S.~R. Depuru, L.~Wang, and V.~Devabhaktuni, ``Smart meters for power
  grid: Challenges, issues, advantages and status,'' \emph{Renewable and
  Sustainable Energy Reviews}, vol.~15, no.~6, pp. 2736 -- 2742, 2011.
\BIBentrySTDinterwordspacing

\bibitem{jacobsen2014}
R.~Jacobsen, N.~T\o{}rring, B.~Danielsen, M.~Hansen, and E.~Pedersen, ``Towards
  an app platform for data concentrators,'' in \emph{Innovative Smart Grid
  Technologies Conference (ISGT), 2014 IEEE PES}, Feb 2014, pp. 1--5.

\bibitem{Weiss2012}
\BIBentryALTinterwordspacing
M.~Weiss, A.~Helfenstein, F.~Mattern, and T.~Staake, ``{Leveraging Smart Meter
  Data to Recognize Home Appliances},'' \emph{IEEE International Conference on
  Pervasive Computing and Communications}.
\BIBentrySTDinterwordspacing

\bibitem{Ruzzelli2010}
\BIBentryALTinterwordspacing
A.~G. Ruzzelli, C.~Nicolas, a.~Schoofs, and G.~M.~P. O'Hare, ``{Real-Time
  Recognition and Profiling of Appliances through a Single Electricity
  Sensor},'' \emph{7th Annual IEEE Communications Society Conference on Sensor,
  Mesh and Ad Hoc Communications and Networks (SECON)}, pp. 1--9, 2010.
\BIBentrySTDinterwordspacing

\bibitem{rune2014}
\BIBentryALTinterwordspacing
R.~Jacobsen and S.~Mikkelsen, ``\BIBforeignlanguage{English}{{Infrastructure
  for Intelligent Automation Services in the Smart Grid}},''
  \emph{\BIBforeignlanguage{English}{Wireless Personal Communications}},
  vol.~76, no.~2, pp. 125--147, 2014.
\BIBentrySTDinterwordspacing

\bibitem{liao2014non}
J.~Liao, G.~Elafoudi, L.~Stankovic, and V.~Stankovic, ``Non-intrusive appliance
  load monitoring using low-resolution smart meter data,'' in \emph{2014 IEEE
  International Conference on Smart Grid Communications (SmartGridComm)}, pp.
  535--540.

\bibitem{Burden:2011}
H.~Burden and R.~Heldal, ``{Natural Language Generation from Class Diagrams},''
  in \emph{Proceedings of the 8th International Workshop on Model-Driven
  Engineering, Verification and Validation}, ser. MoDeVVa.\hskip 1em plus 0.5em
  minus 0.4em\relax ACM, 2011.

\bibitem{witten2005data}
I.~H. Witten and E.~Frank, \emph{Data Mining: Practical machine learning tools
  and techniques}.\hskip 1em plus 0.5em minus 0.4em\relax Morgan Kaufmann,
  2005.

\bibitem{acceleo}
{Object Management Group}, ``{Acceleo},'' {http://www.eclipse.org/acceleo/},
  2015.

\bibitem{Marceau20001389}
\BIBentryALTinterwordspacing
M.~Marceau and R.~Zmeureanu, ``Nonintrusive load disaggregation computer
  program to estimate the energy consumption of major end uses in residential
  buildings,'' \emph{Energy Conversion and Management}, vol.~41, no.~13, pp.
  1389 -- 1403, 2000.
\BIBentrySTDinterwordspacing

\bibitem{reinhardt12tracebase}
\BIBentryALTinterwordspacing
A.~Reinhardt, P.~Baumann, D.~Burgstahler, M.~Hollick, H.~Chonov, M.~Werner, and
  R.~Steinmetz, ``{On the Accuracy of Appliance Identification Based on
  Distributed Load Metering Data},'' pp. 1--9, 2012.
\BIBentrySTDinterwordspacing

\bibitem{plugwise10}
{PlugWise BV, NL}, ``Plugwise circle,'' Online:
  \url{https://www.plugwise.com/circle}, 2015.

\end{thebibliography}

\end{document}